\documentclass[11pt, a4paper]{article}
\usepackage{adjustbox}
\usepackage{srcltx}
\usepackage{amsmath,latexsym,amsfonts,amssymb}
\usepackage{pstricks}
\usepackage[latin1]{inputenc}
\usepackage{epsfig}
\usepackage{fancyhdr}
\usepackage{graphics,graphicx}
\usepackage{xspace}
\usepackage{enumerate}
\usepackage{rotating}
\usepackage{multirow}
\usepackage[linesnumbered,ruled,vlined,,procnumbered]{algorithm2e}
\usepackage{float}
\usepackage{pifont}
\usepackage{tikz}
\usepackage{pgfplots}
\usepackage[round]{natbib}
\usepackage{color}
\usepackage{tablefootnote}
\usepackage{authblk}
\usepackage{amssymb}
\usepackage{url}
\usepackage{hyperref}
\usepackage{longtable}
\usepackage{comment}

\usepackage{subcaption}
\usepackage{makecell}

\makeatletter
\AtBeginEnvironment{function}{%
	\let\c@algocf\c@function
}
\makeatother

\begin{document}
	
	\title{Strategic energy flows in input-output relations: a temporal multilayer approach}
	
	\author[(1)]{Gian Paolo Clemente}
	\author[(2)]{Alessandra Cornaro}
	\author[(2)]{Rosanna Grassi}
	\author[(2)]{Giorgio Rizzini}

	\affil[(1)]{\small{Department of Mathematics for Economic, Financial and Actuarial Sciences \\ Università  Cattolica del Sacro Cuore, Milan, Italy}}
		
	\affil[(2)]{\small{Department of Statistics and Quantitative Methods\\ University of Milano - Bicocca, Via Bicocca degli Arcimboldi 8, 20126 Milano, Italy}}
		
		\date{}

		\maketitle
		
	\begin{abstract} 
	The energy consumption, the transfer of resources through the international trade, the transition towards renewable energies and the environmental sustainability appear as key drivers in order to evaluate the resilience of the energy systems. Concerning the consumptions, in the literature a great attention has been paid to direct energy, but the production of goods and services also involves indirect energy.
	Hence, in this work we consider different types of embodied energy sources and the time evolution of the sectors' and countries' interactions. Flows are indeed used to construct a directed and weighted temporal multilayer network based respectively on renewable and non-renewable sources, where sectors are nodes and layers are countries. We provide a methodological approach for analysing the network reliability and resilience and for identifying critical sectors and economies in the system by applying the Multi-Dimensional HITS algorithm. Then, we evaluate central arcs in the network at each time period by proposing a novel topological indicator based on the maximum flow problem. In this way, we provide a full view of economies, sectors and connections that play a relevant role over time in the network and whose removal could heavily affect the stability of the system. 
	We provide a numerical analysis based on the embodied energy flows among countries and sectors in the period from 1990 to 2016. Results prove that the methods are effective in catching the different patterns between renewable and non-renewable energy sources.
\end{abstract}

\textbf{Keywords:} Temporal Multilayer networks, Embodied energy, Multi-dimensional HITS, Energy security, Max flow problem

\section{Introduction}
\label{intro}
Modern energy systems are becoming essential for the provision of several primary services, such as for instance clean water, sanitation and healthcare, reliable and efficient lighting, heating, cooking, mechanical power, transport and telecommunications services, and they have led to significant improvements in the standard of living for part of the world's population. The importance of energy is well stressed in a famous speech (see \cite{Schumacher})
where it has been defined as "not just another commodity, but the precondition of all commodities, a basic factor equal with air, water, and earth".
Typically an energy system refers to a wide complex structure that involves the set of elements including production, transformation, transport and distribution of energy sources. Nowadays the attention to the different phases of the process is increased because of several aspects. \\
First, research is focusing on energy security as a key variable in the relationship between energy and well-being (see, e.g., \cite{Hern}). In particular, great attention is paid to the possibility of assuring an uninterrupted availability of energy sources at an affordable price.
Energy security involves two facets. 
In a long-term view, it mainly deals with investments to supply energy in line with economic developments and environmental needs. Short-term energy security focuses instead on the ability of the energy system to be resilient and react promptly to sudden changes in the supply-demand balance. \\
Second, oil, coal and natural gas still meet the surging world energy demand, creating serious implications for the public health (see \cite{Fritz}) and the environment. One of the results is that carbon dioxide emissions, the principal cause of global warming, are rising and hit a new record in 2022. Therefore, a close link between efforts to ensure energy security and those to mitigate climate change becomes crucial. Indeed, as pointed out in (\cite{iea}), to optimise the efficiency of their energy policy, countries must consider energy security and climate change mitigation priorities jointly. \\
Third, energy access is also an area of great inequity. Energy sources are indeed not equitably distributed around the globe creating an asymmetric world with a strong dependence on major fossil fuel exporters (see, e.g., \cite{Overland}, \cite{Overland2} and \cite{Van}) and favouring the energy poverty in developing countries (see \cite{Sy} and \cite{Thomson}).\\
In this context, the energy consumption, the transfer of direct and indirect energy resources through the international trade, the transition towards renewable energies, and the environmental sustainability appear as key drivers in order to evaluate both the resilience of the energy systems and to address the decisions of policymakers towards sustainable policies.\\ 
Focusing on the energy consumption, a common practice is to pay attention to direct energy. However, the production of goods and services 
also involves indirect energy.
This one (also called embodied energy) is the energy consumption embodied in goods and services production that is needed to make the final product (see, e.g., \cite{Brown} and \cite{qier}).
Hence, to catch completely the complexity and the wide range of energy transfers among the economies, we provide a framework based on the global, direct and indirect, energy involved in the production of goods and services.

Energy embodied in international trade has been intensively analysed in the literature (see \cite{Miller}, \cite{Sun} and \cite{Wiedmann2}). In particular, interdependencies between industries within the economy has been initially depicted by Leontief (see \cite{Leontief1} and \cite{Leontief2}) that shows how output from one industrial sector may become an input to another one. 
An extended input-output analysis (EEIOA) (see, among others, \cite{Kites}, \cite{Miller} and \cite{Wiedmann}) has been further developed to evaluate the environmental and social impact of the economic activities, including the harvest and degradation of natural resources. In this field, the interconnected and complex embodied energy flows in international trade have been studied from a network perspective. To this end, many suitable tools have been applied to consider several direct energy commodity trade networks and to explore the characteristics of the embodied energy in trade at global, national and regional levels (see, e.g., \cite{Chen2}, \cite{Du}, \cite{Hao} and \cite{Zhong}). Typically this topic is modelled in the literature by using monoplex networks. However, this representation does not capture the joint interaction between sectors and countries, being based on an analysis at either country or sector level. The proposed approaches have been extended in \cite{RizCor}, where a multilayer network has been considered. 

In this work we extend the approach of \cite{RizCor}, considering
different types of embodied energy sources as well as the time evolution of the sectors' and countries' interactions.  Therefore, we focus on 
the embodied energy flow on temporal multilayer networks based respectively on renewable and non-renewable sources. Hence, we extend existing methodologies provided in \cite{Chen2} and \cite{RizCor}, by computing the embodied energy flows for each couple of countries and sectors in each time period and for different energy sources. Given a specific time and a specific energy source, flows are used to construct a directed and weighted multilayer network where sectors/industries are nodes and each layer represents the economy/country\footnote{Notice that, in what follows, we use industries and sectors as synonyms. Moreover, economies and countries are also used as synonyms.}. Weighted intra-layer arcs consider the entity of the directed embodied energy flows, based on either renewable or non-renewable sources, between different sectors of the same economy in a specific time period. Similarly, weighted inter-layers arcs describe directed flows between
the same (or different) sector in different countries . By combining the multilayer networks at each time periods for the same energy sources (renewable or non-renewable), we can build a temporal multilayer network that allows to catch both the interactions between sectors and countries and the evolution over time. \\
Then, we provide a methodological approach based on a temporal multilayer network for analysing the network reliability and resilience and for identifying critical sectors and economies in the system. As stressed before, enhancing resilience of energy systems is becoming a crucial issue nowadays (see, e.g., \cite{Sang}) and the identification of relevant players in the market can be helpful.  To this end, we use a Multi-Dimensional HITS (MD-HITS), developed by \cite{Arrigo}, to identify economies and sectors that play a key role acting as collectors and distributors in global energy flow system. Additionally, 
we evaluate central arcs in the network at each time period by using a method based on the maximum flow problem (see \cite{fordfulkerson}).
In this way, we provide a full view over time of economies, sectors and connections that play a relevant role over time in the network and whose removal could heavily affect the stability of the system. The proposed measures are new in this context, 
since embodied energy flows have been either studied using classical centrality measures based on monoplex networks (see \cite{Chen2}, \cite{Du} and \cite{Hao}) or focusing only on nodes and layers in a given period in a multilayer framework (see \cite{RizCor}). \\
The identification of strategic sectors can also represent a useful tool for the policymakers in order to develop policies that consider both the economic relevance and the environmental impact of different industries and to assure that the challenge to decarbonize for sectors, that produce the majority of global greenhouse-gas emissions, can be managed considering also a sustainable economy growth. \\
A detailed numerical case study has been developed based on the data provided by the multi-region input-output tables\footnote{Data are taken from the the Eora Global Supply Chain Database. See https://worldmrio.com/}. In particular, we constructed a temporal multilayer network based on the embodied energy flow among countries and sectors in the period from 1990 to 2016. In each year, we have a directed and weighted multilayer network with 189 layers, given by the countries, and 26 nodes given by the sectors. Then, the proposed methods have been applied in order to detect central nodes and layers in the network both in the whole period and in each year. Additionally, relevant arcs have been identified in each period. The analysis has been repeated considering three different framework based on the use of all energy source, only renewable ones or only non-renewable ones. To this end, we are also able to emphasize countries that could emerge as key players during the transition from non-renewable to renewable energies. \\
Results related to the global energy are in line with \cite{Chen2} and \cite{RizCor} where relevant countries in the world trade have also a central role in terms of consumption. However, additional insights are provided. In particular, an interesting behaviour is represented by the evolution of countries' ranking over time with respect to different energy sources. Along with main economies in the world, we observe the increasing role of Middle East when non-renewable sources are considered. On the other hand, Nordic countries are confirmed in the top position when green sources are considered. Focusing on strategic arcs, we notice that the development of the economy and the geographical proximity appear two relevant factors. Connections involving United States with Canada and other relevant players in the Asia-Pacific area have indeed a relevant role in the network.  However, when renewable energy is considered, it is noticeable the presence of countries that  are at advanced stages of the transition and well positioned to export green energy.
Results confirm that the articulated structure of the temporal multilayer network can provide several indications useful for assessing the resilience of the system and for identifying critical players. \\
The paper is organized as follows. In Section \ref{sec:Pre}, we describe preliminaries and the mathematical notations. Section \ref{sec:meth} provides the methodology used for building the multilayer network as well a description of the Multi-Dimensional Hits approach applied for assessing nodes, layers and time centrality scores. We also propose a novel method for identifying the centrality of arcs based on the evaluation on the total max flow of the removal of the arc. In Section \ref{sec:na}, we report the numerical analysis. Conclusions follow. Maximum flow formulation problem, as well as the list of sectors and countries are reported in the Appendices.

\section{Preliminaries}
\label{sec:Pre}

In this section we recall some notations and definitions about directed graphs (more detailed definitions can be found, for instance, in \cite{Bang-Jensen2008}).\\ 
A directed and weighted network is defined by a graph 
$G=(V,E,w)$, where $V$ is the set of $N$ vertices (or nodes), $E$ is a subset of $V\times V$ 
(set of arcs or directed edges) and $w$ is a real positive weight assigned to each arc. 
Two nodes $i,j \in V$ are adjacent if there is at least an arc $(i,j)$ starting from $i$ and ending in $j$. 
The graph $G=(V,E,w)$ can be described by a real $N$-square matrix $\mathbf{W}^{[\alpha]}$ (the weighted adjacency matrix) whose entries are $w^{[\alpha]}_{ij}>0$ if $(i,j)\in E$, and $ 0$ otherwise.\footnote{For the sake of clarity, we use the notation $\textbf{W}^{[\alpha]}$ for the weighted adjacency matrix, to be conform with the  notation of the weighted supradjacency matrix adopted for multilayer networks.}
Notice that the entry $w_{ij}^{[\alpha]}$ represents the weighted arc from node $i$ to node $j$. In other words, the $i-th$ row encloses all the weighted arcs outgoing from $i$. 

If matrix $\mathbf W^{[\alpha]}$ is irreducible (this is guaranteed for strongly connected graphs), the spectral radius $\rho >0$ is a simple eigenvalue, associated with the right eigenvector $\mathbf x$ whose components are all positive (see \cite{Horn2012}). This definition is meaningful in view of the definition of vertex centrality we will introduce.
Centrality is one of the essential issues in network theory. In this work, we focus on the eigenvector centrality. For a node $i$, this measure is defined as the $i$-th component of the eigenvector associated with the spectral radius $\rho$ of the weighted adjacency matrix (see \cite{Bonacichb} and \cite{Bonacicha}). In matrix representation:

\begin{equation}\label{eigen}
	\textbf{x}^{[\alpha]}=\frac{1}{\rho}{\textbf{W}^{[\alpha]}}\textbf{x}^{[\alpha]},
\end{equation}
where $\textbf{x}^{[\alpha]}$ is the vector collecting the centrality scores of the nodes.\\
The measure well represent the power position of a vertex in the whole network: by definition, the centrality of a vertex  $x^{[\alpha]}_i$ depends on the centralities of its neighbours, then can be amplified when a vertex is connected to nodes which are central themselves.

The measure can be extended to directed networks, by defining two centrality measures, hub and authority, associated with a node (see \cite{Kleinberg1999}). Let $x^{[\alpha]}_i$ and $y^{[\alpha]}_i$ be the hub and authority scores for the node $i$, respectively. The centrality values for all nodes are the components of two hub and authority vectors 
$\textbf{x}^{[\alpha]}$ and $\textbf{y}^{[\alpha]}$. 

These vectors are obtained as the limit values by the following recursive equations when $k \to \infty$: 
\begin{equation}\label{auth}
	\textbf{y}^{[\alpha]}_{(k)}={\textbf{W}^{[\alpha]}}^T\textbf{x}^{[\alpha]}_{(k-1)}
\end{equation}
and 
\begin{equation}\label{hubs}
	\textbf{x}^{[\alpha]}_{(k)}=\textbf{W}^{[\alpha]}\textbf{y}^{[\alpha]}_{(k)}.
\end{equation}

where $\textbf{x}^{[\alpha]}_{(k)}$ and $\textbf{y}^{[\alpha]}_{(k)}$ collect the hub and authority scores, respectively, of all nodes at the $k-th$ step.
By (\ref{auth}) and (\ref{hubs}), a mutually reinforcing relationship characterizes hubs and authorities: a high hub actor points to many good authorities, and a high authority actor receives from many good hubs.
Equations (\ref{hubs}) and (\ref{auth}) are the basis of the iterative algorithm (HITS - Hyperlink Induced Topic Search) proposed by \cite{Kleinberg1999}.

We now introduce the definitions and the adopted notations of multilayer networks.
A multilayer
network is a family of networks $G_\alpha=(V_\alpha,E_{\alpha,\beta})$,  $\alpha,\beta=1,...,L$, where each network $G_\alpha=(V_\alpha,E_{\alpha,\beta})$ is located in a layer $\alpha$ and a node $i \in V_\alpha$ is adjacent to $j \in V_\beta$, $\forall \alpha,\beta=1,...,L$ if there is an arc connecting them. \\
In this work we consider node-aligned networks with non-diagonal couplings. 
In node-aligned networks, nodes are the same over all layers, namely, $V_\alpha=V_\beta=V,\ \forall \alpha,\beta=1,...,L$, and
$E_{\alpha,\beta}$ collects all the arcs connecting nodes on layer $\alpha$ to nodes within the same layer ($\alpha=\beta$, intra-layer connections) and to nodes on different layers ($\alpha \ne\beta$, inter-layer connections). If non-diagonal couplings are allowed, inter-layer arcs can exist also between different nodes in different layers. We refer the reader to \cite{Kivela} for a meaningful classification of multidimensional networks.

\noindent We associate a weight $w^{\left[\alpha\beta\right]}_{ij} > 0$ with an arc $(i,j)$ in $E_{\alpha \beta}$ and,
when $\alpha=\beta$, we intend that there is a weighted arc $w^{\left[\alpha\right]}_{ij} > 0$ between nodes $i$ and $j$ in the network $G_\alpha$.

As for the monoplex case, a matrix describes the adjacency relationships. In the multilayer framework, the weighted supradjacency matrix is defined as a $L-$square blocks matrix, with each block of order $N$:

\begin{equation}
	\label{supra}
	\mathbf{W}=
	\begin{bmatrix}
		\mathbf{W}^{\left[1\right]} & \mathbf{W}^{\left[12\right]} & \cdots & \mathbf{W}^{\left[1L\right]} \\
		\mathbf{W}^{\left[21\right]} & \mathbf{W}^{\left[2\right]} & \cdots & \mathbf{W}^{\left[2L\right]} \\
		\vdots & \vdots & \ddots & \vdots \\
		\mathbf{W}^{\left[L1\right]} & \mathbf{W}^{\left[L2\right]} & \cdots & \mathbf{W}^{\left[L\right]}
	\end{bmatrix},
\end{equation}

\noindent where the diagonal blocks represent the weighted adjacency matrix of each layer $\textbf{W}^{\left[\alpha \right]}$, $\alpha=1,...,L$, whereas the non-diagonal blocks $\textbf{W}^{\left[\alpha \beta\right]}$,  $\alpha,\beta=1,...,L$ represent the weighted adjacency relations between nodes on layers $\alpha$ and nodes on layer $\beta$. 

\noindent We now denote as  $w_{hk}$, the generic element of the matrix $	\mathbf{W}$ with $h,k=1,...,NL$, where:
\begin{equation}
	h=N(\alpha-1)+i, k=N(\beta-1)+j.
	\label{eq:hk}
\end{equation} 
Formula (\ref{eq:hk}) relates the indices $h,k$ to the position in the supradjacency matrix $\mathbf{W}$ of the weight of the arc $(i,j) \in E_{\alpha,\beta}$ (i.e. $w^{\left[\alpha\beta\right]}_{ij}=w_{hk}$). For the sake of simplicity, throughout the paper, we sometimes use the notation $w_{hk}$ instead of the complete notation.
Thus, from now on, we always assume relation (\ref{eq:hk}) between $h,k$ and $\alpha,\beta,i,j$.\\
We now introduce the concept of temporal network in the multilayer context.
A temporal network (see \cite{Holme}), can be modelled in terms of time-varying graphs, which are time-ordered sequences of graphs over a set of nodes. Hence, a time-stamp is also included in the edge''s tuple. As in \cite{Holme}, we assume that a set of vertices interacts with each other at certain times, and the durations of the interactions are negligible. This allows us to represent the temporal multilayer network $\mathcal{G}$ as a sequence of networks $\mathcal{G}_{t}$ where $t$ belongs to a finite discrete time horizon, i.e. $t \in \{1,2,...,T\}$, where $T$ is the number of instants.
Networks $\mathcal{G}_{t}$ are defined over the same set of nodes and layers and the arcs describe, at each time, the existing connections. Hence, a time-dependent weight $w^{\left[\alpha\beta\right]}_{ij}(t)$ is associated with each arc present at time $t$ in the network. 

\section{Methodology and data}
\label{sec:meth}
\subsection{Global embodied energy flow in a temporal multilayer network}
\label{sec:n1}

In this section we describe the methodology used to define the global embodied energy flow for a temporal multilayer network $\mathcal{G}$.\\
In particular, we consider a directed and weighted 
temporal multilayer network with $N$ nodes and $L$ layers, representing sectors and economies, respectively. 
To facilitate the reading, we will indicate the sectors by the Latin letters $i,j$ and the economies by the Greek letters $\alpha$ and $\beta$,  and we will adopt the relation (\ref{eq:hk}) between the indices $h,k$ and $i,j,\alpha,\beta$ wherever possible. 
At each time $t$, the flows between different sectors within the same economy are represented by intra-layer connections, while inter-layers take into account the flows between the same or different sectors in different economies. 
Therefore, nodes are connected by weighted arcs, where the time-dependent weights $w^{\left[\alpha\beta\right]}_{ij}(t)$ describe the global embodied energy flow between pairs of sectors and economies at time $t$.
A weighted supradjacency matrix $\mathbf{W}_{t}$, of order $NL$, defined as in formula (\ref{supra}), fully describes the multilayer network $\mathcal{G}_{t}$ at each time $t$. In particular, $\mathbf{W}_{t}$ is a block matrix where the blocks that belong to the main diagonal represent the embodied energy flows between sectors related to the same economy, whereas the non-diagonal blocks are referred to embodied energy flows between sectors of different economies. \\
The entries $w_{hk}(t)$ of the matrix  $\mathbf{W}_{t}$ 
are computed following the environmentally extended input-output analysis (see, among others, \cite{Kites}, \cite{Miller} and \cite{Wiedmann}).\\
Let us consider, at each time $t$, a pair of sectors $i,j$, with $i,j=1, \dots, N$ and a pair of economies  $\alpha, \beta$, with $\alpha, \beta=1, \dots, L$, respectively.\\
First of all we need the so-called Leontief inverse matrix in order to estimate the technical energy consumption of all considered sectors assigned to the final use.
The Leontief inverse matrix of order $NL$ is defined as:
\begin{equation}
	\mathbf{L}_{t}=(\mathbf{I}-\mathbf{A}_{t})^{-1},
\end{equation}
where  $\mathbf{I}$ is the identity matrix and $\mathbf{A}_{t}$ is the input coefficient matrix at time $t$.
The entries $a_{hk}(t)$ of $\mathbf{A}_{t}$ are given by:

\begin{equation}
	a_{hk}(t)=\dfrac{u_{ij}^{\alpha \beta}(t)}{o_{j}^{\beta}(t)},
\end{equation}
where $u_{ij}^{\alpha \beta}(t)$ is the intermediate use of sector $j$ in economy $\beta$ provided by sector $i$ in economy $\alpha$ at time $t$ and $o_{j}^{\beta}(t)$ is the total output of sector $j$ in economy $\beta$ at time $t$. Notice that the previous formula stands in view of the relation defined in (\ref{eq:hk}).

Now, the embodied energy flow can be defined as:
\begin{equation}
	\label{qij_network}
	q_{ij}^{\alpha \beta}(t)= \left( \sum_{\epsilon=1}^{L}  c_{i}^{\epsilon}(t) l_{ij}^{\epsilon\alpha}(t) \right)  d_{j}^{\alpha \beta}(t) ,
\end{equation}

where:
\begin{itemize}
	\item $c_{i}^{\epsilon}(t)$, with $\epsilon=1, \dots, L$, stands for the technical energy consumption by sector $i$ in each investigated economy $\epsilon$ at time $t$; 
	
	\item $l_{ij}^{\epsilon\alpha}(t)$,  with $\epsilon=1, \dots, L$, is the entry of the Leontief inverse matrix  $\mathbf{L}_{t}$ for sectors $i$ and $j$ and economies $\epsilon$ and $\alpha$ at time $t$;
	
	\item $d_{j}^{\alpha \beta}(t)$ represents the goods/services traded as a final demand by sector $j$ from economy $\alpha$ to economy $\beta$ at time $t$.
	
\end{itemize}

Value of $q_{ij}^{\alpha \beta}(t)$ stands for the total embodied energy that is needed from $i$ in order to produce the good/services that are traded in sector $j$ as a final demand from economy $\alpha$ to economy $\beta$ at time $t$.
We define $w_{hk}(t)=q_{ij}^{\alpha \beta}(t)$ if $q_{ij}^{\alpha \beta}(t)>0$ and $0$ otherwise. 

\subsection{MD-HITS and node and layer centrality}
\label{sec:n2}

We now recall a generalization of the Kleinberg algorithm for hub and authority scores 
called Multi-Dimensional HITS (MD-HITS), proposed by \cite{Arrigo}. 
MD-HITS accounts for layers and time stamps as dimensions with the aim to compute node and layer centrality measures. 
The model in \cite{Arrigo} defines five centrality vectors: two for nodes (hub and authority scores), two for layers (broadcasting and receiving scores) and one for the time dimension (see Definition 1 in \cite{Arrigo}).

In our framework, this approach allows us to identify which are the economies and sectors acting as collectors and distributors in the 
analysed temporal multilayer network. In doing so, we consider different types of energies: renewables, non-renewables and the total energy.
In particular, we consider two vectors of centrality for nodes: a  vector $\mathbf{x}$ of components $x_i$, with $i=1, \dots, N$, of hub scores and a vector $\mathbf{y}$ of components $y_j$, with $j=1, \dots, N$ of authority scores.
Similarly, we have two vectors, $\mathbf{b}$ of components $b_\alpha$ and $\mathbf{z}$ of components $z_\beta$, with $\alpha,\beta=1, \dots, L$,  that account for the broadcasting and receiving capability of layers, respectively. Furthermore we introduce a vector  $\mathbf{u}$ of elements $u_t$ with $t \in \{1,...,T\}$ that accounts for time scores. \\
The centrality values are computed by using the following relations (see Algorithm 1 in \cite{Arrigo}):

\begin{align*}
	x_{i} = \sum_{j,\alpha,\beta, t} \left( w_{ij}^{[\alpha \beta]}(t)  y_{j}  b_{\alpha}  z_{\beta} u_{t}\right) ^{\gamma_1} , \\
	y_{j} = \sum_{i,\alpha,\beta,t} \left( w_{ij}^{[\alpha \beta]}(t)  x_{i}  b_{\alpha}  z_{\beta} u_{t}\right) ^{\gamma_2}, \\
	b_{\alpha} = \sum_{i,j,\beta,t} \left( w_{ij}^{[\alpha \beta]}(t)  x_{i}  y_{j} z_{\beta} u_{t}\right) ^{\gamma_3}, \\
	z_{\beta} = \sum_{i,j,\alpha,t} \left( w_{ij}^{[\alpha \beta]}(t)  x_{i}  y_{j}  b_{\alpha} u_{t}\right) ^{\gamma_4}, \\
	u_{t} = \sum_{i,j,\alpha,\beta} \left( w_{ij}^{[\alpha \beta]}(t)  x_{i}  y_{j}  b_{\alpha} 	z_{\beta}\right) ^{\gamma_5},
\end{align*}
with $\gamma_{k} \in (0,1]$, with $k=1,...,5$.
It is noteworthy to observe that the vector $\boldsymbol{\gamma}=(\gamma_{1},...,\gamma_{5})$ assures that the centralities previously introduced are well defined.
Indeed, following \cite{Arrigo}, in absence of information about the entries of $\boldsymbol{\gamma}$, a uniform choice can be inferred and, in this specific case, we set $\gamma_k=\frac{1}{5}$, with $k=1,...,5$.
This guarantees the existence and uniqueness of the proposed centrality measures in virtue of Theorem 2.1 of \cite{Arrigo}.

\subsection{Max Flow and arc centrality}
\label{sec:n3}
In this section, we propose a new index to assess the relevance of an arc in a multilayer network. The methodology is based on the solution of the max flow problem\footnote{For a more detailed treatment of the problem, we refer the reader to \cite{kleinberg2006}.}, which is briefly described in the Appendix \ref{appsec:maxflow}.\\
Given the multilayer $\mathcal{G}_t$ at time $t$, 
we calculate the max flow between any pair of nodes $i$ and $j$, where $i$ lies on layer $\alpha$ and $j$ on layer $\beta$, and we collect them in a $NL$ - square matrix $\mathbf{F}_t$ of entries $f_{hk}(t)=f^{\left[\alpha \beta\right]}_{ij}(t)$. We define the following quantity:
\begin{equation}
	\bar{m}_{t} = \sum_{h,k}{f}_{hk}(t)
	\label{barM}
\end{equation}
which expresses the overall maximum flow circulating in the multilayer at time $t$. 
In order to assess the relevance (in terms of flows) of the arcs in $\mathcal{G}_t$, we construct a new multilayer network $\widetilde{\mathcal{G}}_{t}$, obtained by deleting an arc in $\mathcal{G}_t$.\\
The maximum flow problem is then solved for all couple of nodes $i$ on layer $\alpha$ and $j$ on layer $\beta$ of the multilayer network $\widetilde{\mathcal{G}}_{t}$ and the results are collected in a new matrix\footnote{We notice that the procedure is repeated for all the arcs of $\mathcal{G}_t$. Therefore, the number of generated matrices $\widetilde{\mathbf{F}}_t$ is, at the end, equal to the number of arcs of $\mathcal{G}_t$.} $\widetilde{\mathbf{F}}_t$. 
The overall maximum flow circulating in the multilayer $\widetilde{\mathcal{G}}_t$ at time $t$ is now defined as:
\begin{equation}
	\widetilde{m}_{t} = \sum_{h,k} \widetilde{f}_{hk}(t),
	\label{tildeM}
\end{equation}
where $\widetilde{m}_{t}$ expresses the total maximum flow of $\widetilde{\mathcal{G}}_t$ once a generic arc has been eliminated. \\
The relevance of the deleted arc at time $t$ can be modelled through the following index:
\begin{equation}
	\widetilde{r}_t= 1- \frac{\widetilde{m}_{t}}{\bar{m}_{t}}.
	\label{coeff_r}
\end{equation}
It is easy to observe that the index $\widetilde{r}_t$ in \eqref{coeff_r} lies in the interval $[0,1)$\footnote{The proof is trivial by using simple linear algebra results.}. 

A null value for $\widetilde{r}_t$ means that the coefficients $\widetilde{m}_{t}$ and $\bar{m}_{t}$ are equal. In this case the deletion of an arc does not affect the overall maximum flow, i.e. the maximum flow that can circulate in the multilayer networks $\widetilde{\mathcal{G}}_t$ and $\mathcal{G}_t$ is the same. 
Hence, the index $\widetilde{r}_t$ reflects the relevance of the considered arc in the multilayer. 
Indeed, if the deletion of the arc leads to a decrease of the total max flow in $\widetilde{\mathcal{G}}_t$, the ratio $\frac{\widetilde{m}_{t}}{\bar{m}_{t}}$ decreases and, consequently, $\widetilde{r}_t$ increases.


\section{Numerical application}
\label{sec:na}
We consider here a temporal multilayer network $\mathcal{G}$ based on the embodied energy flow among economies and sectors in the period ranging from the beginning of 1990 to the end of 2016. The list of sectors and countries and the related codes are reported in the Appendix \ref{sectorscountrieslists} (Tables \ref{listsectors} and \ref{tab:tableofcountries_1}, respectively). Data are taken from EORA Global Supply Chain Database that consists in a multi-region input-output table 
documenting the inter-sectoral transfers across countries (see \cite{lenzen2013}). \\
Following the definition of temporal multilayer networks given in Section \ref{sec:Pre}, we summarise data of each year in a different network $\mathcal{G}_{t}$, with $t=1,...,27$. 
Each multilayer network $\mathcal{G}_{t}$, related to the instant $t$, is characterised by $189$ layers, representing the economies, and $26$ nodes, associated to the different sectors. Nodes are connected by weighted arcs describing the embodied energy flows specified in formula \eqref{qij_network}.  Based on \cite{Chen2} and \cite{RizCor}, to calibrate the weights, we have considered six final-demand coupled system provided by EORA Global MRIO. Additionally, the approach has been applied separately three times in order to consider in a different way the alternative energy sources. Therefore, for each time period, we have $\mathcal{G}^{A}_{t}$,  $\mathcal{G}^{R}_{t}$, $\mathcal{G}^{NR}_{t}$ that consider the total energy flows, only the renewable sources (biomass and waste electricity, hydroelectricity and other renewable like geothermal, solar, tide, wave and wind) or only the not renewable sources (coal, natural gas, petroleum, nuclear electricity), respectively. \\

The first part of the section focuses on the original input-output relations among sectors from different countries. In particular, main patterns over time in terms of consumes are described so that most relevant sectors and countries are considered. The rest of the section is devoted to the identification of central nodes and edges, respectively. In particular, with respect to different times and sources of energies, we are able to identify key sectors and countries and relevant connections in the networks. \\

\subsection{Consumption}
\label{consumption_section}

\begin{figure}[!h]
	\includegraphics[scale=0.23]{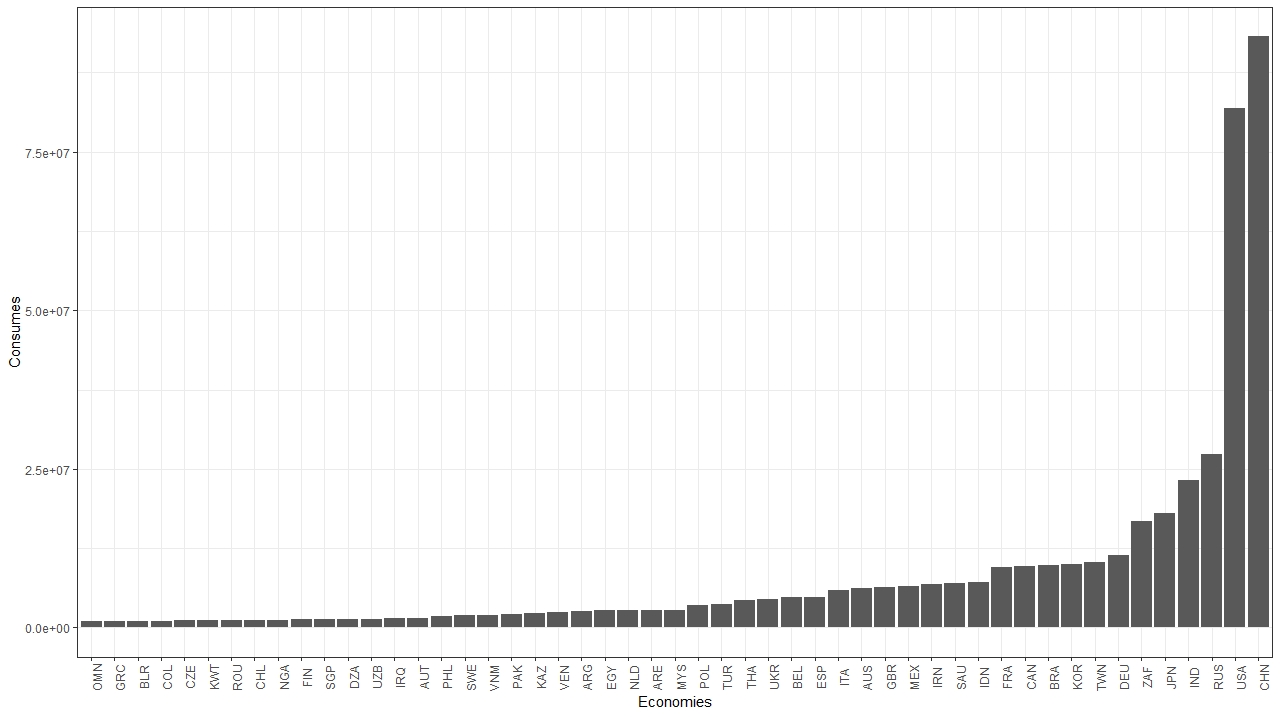}
	\caption{Top 50 economies in terms of energy consumes in 2016.}
	\label{cons_country}
\end{figure}

\begin{figure}[!h]
	\includegraphics[scale=0.25]{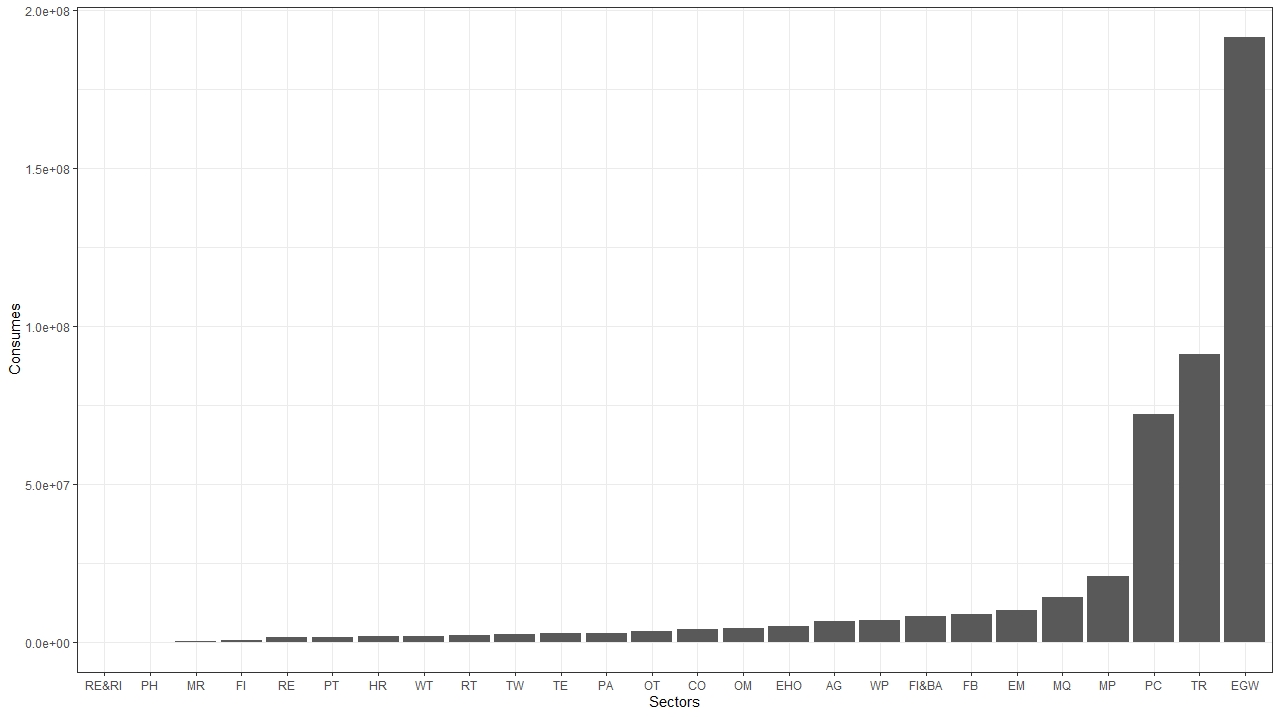}
	\caption{Energy consumption of sectors in 2016}
	\label{cons_sector}
\end{figure}

We start in this section to provide some preliminary analyses about energy consumption. In particular, we start focusing on the whole energy consumption over the last observed year (i.e. 2016). To this end, Figures \ref{cons_country} and \ref{cons_sector} report  the sectors and countries characterised by the higher energy consumes. As in \cite{WU2016} and \cite{RizCor}, we notice that sectors characterised by elaborate end-products, that require several processing steps and inputs, belong to the top ranking (see Electricity, Gas and Water (EGW), Petroleum, Chemical and non-Metallic Mineral Products (PC), Transport (TR) and Metal Products (MP)). Most of the industries in these manufacturing sectors are indeed considered to be energy-intensive (see \cite{EIA}). As regards to countries, the main consumers are China, United States, Russia, India and Japan. These results are in line with the data about global electric energy consumption published  in \cite{EIA}, Ernadata\footnote{\url{https://yearbook.enerdata.net/total-energy/world-consumption-statistics.html}} and with the findings in \cite{Chen2} and \cite{RizCor}.\\

\begin{figure}[!h]
	\includegraphics[scale=0.25]{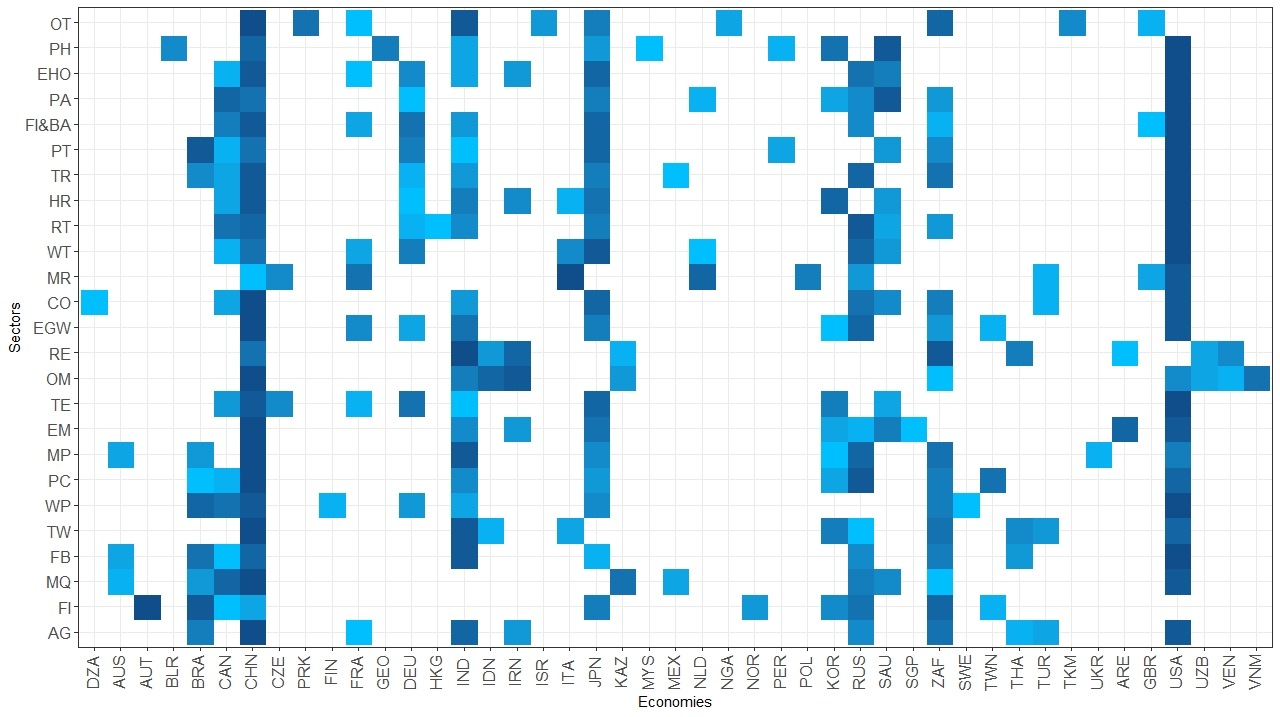}
	\caption{Top 10 countries in terms of energy consumes for each sector in 2016. Darker blue means higher consumes.}
	\label{important_edges}
\end{figure}

As regards to the combination sector-economy, we display in
Figure \ref{important_edges} the top ten countries in terms of energy consumption for each sector. In particular, it is clear the prominent role of China and United States in all the sectors. Moreover, we observe that China is the biggest energy consumer in 10 sectors and belongs to the top five in 23 sectors (see, e.g., \cite{shi2017} for a similar result). Furthermore, 28 out of 30 arcs with higher weights are intra-layer connections in the layer that describes China economy. \\
United States is instead the top consumer in 12 sectors and in the top 5 in 21 sectors.\\
It is also interesting to observe how other relevant players, as Russia, Japan, India, South Africa, appear mainly exposed to sectors characterised by significant consumptions. \\
Looking at the European countries, we notice higher consumes in specific sectors related to the specificity of their economies. For instance, Austria is the biggest energy consumer in Fishing (FI) while Italy in Maintenance and Repair (MR). Because of its morphological structure and its lack of access to the sea, Austria depends largely on imports of marine and freshwater fish. The high energy consumption in fishing sector is explained noticing that, on the one hand, the fish demand is increased in the last years (see \cite{acquaculture}) and, on the other hand, the fish market supply chain requires energy costs such as fuel input, transportation, materials, equipment and services (see \cite{huisingh2015}). For Italy, this evidence can be partially explained by the relevance of Maintenance and Repair industries in particular sectors (e.g. automotive). 

\begin{figure}[!h]
	\begin{subfigure}[t]{1\textwidth}
		\includegraphics[scale=0.25]{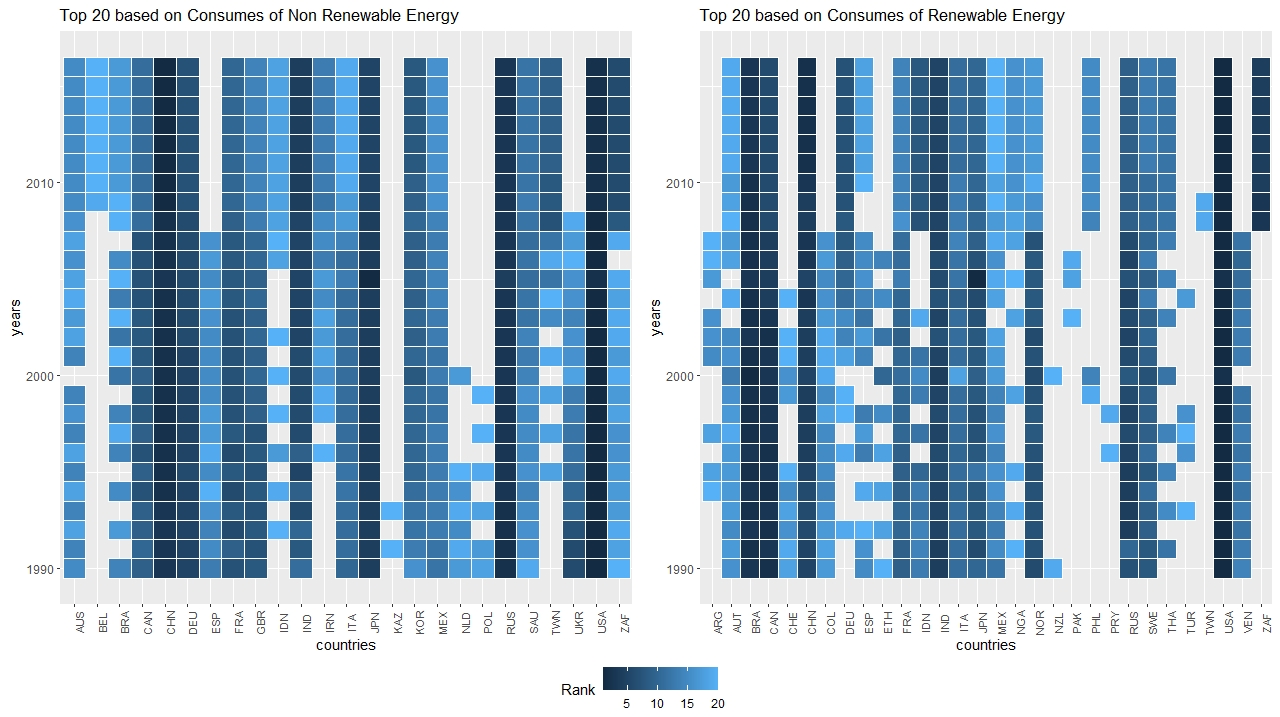}
		\caption{Ranking of countries that belong to the top 20 in terms of energy consumes. We display only countries that belong to the top 20 in at least one year.}
		\label{cons_coutype}
	\end{subfigure} \\
	\begin{subfigure}[t]{1\textwidth}
		\includegraphics[scale=0.25]{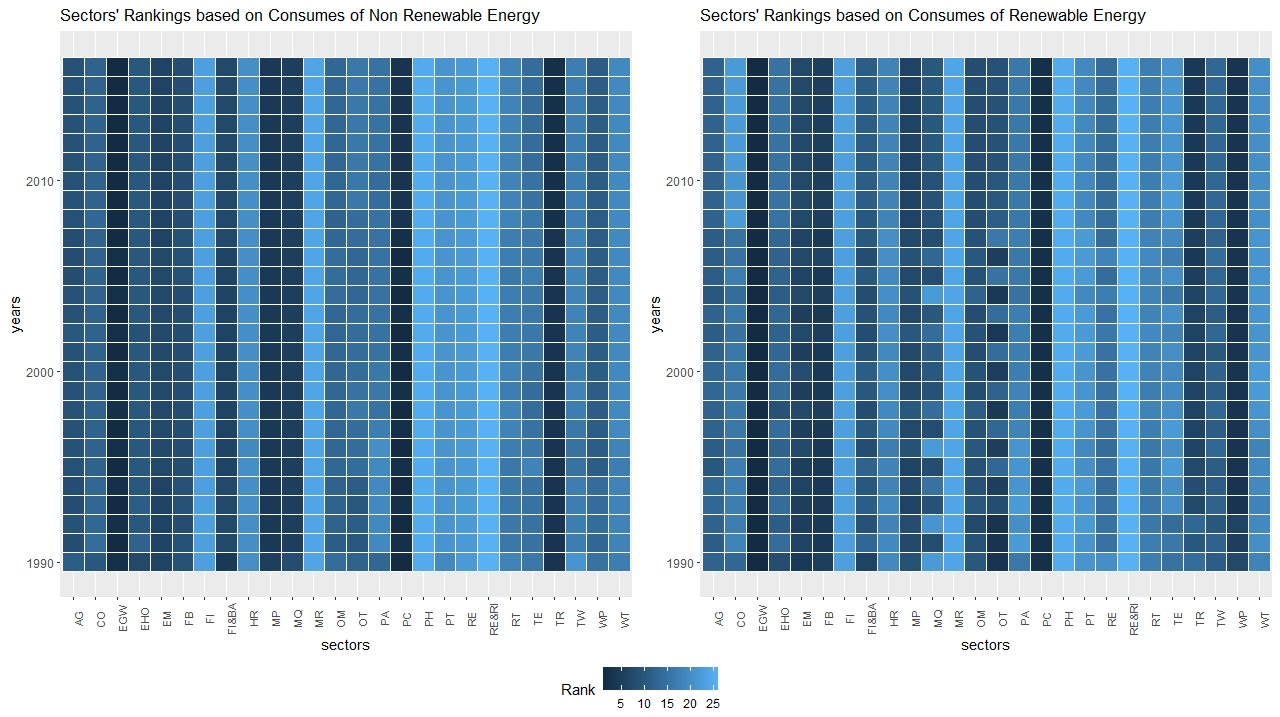}
		\caption{Ranking fo sectors with respect to year and type of energy}
		\label{cons_sectortype}
	\end{subfigure} 
	\caption{Energy consumption of sectors and economies with respect to year and type of energy.}
	\label{cons_totyear}
\end{figure}

Focusing on the temporal evolution, energy consumes increased in the analysed period. In particular from 1990 to 2016, we observe a general increase of 30\% of non-renewable sources and of 45\% of renewable energy. Indeed, demand for energy is growing across many countries in the world as people get richer and populations increase. However, the increased demand is not offset by improvements in energy efficiency, then the global energy consumption has increased nearly every year for more than half a century. Few exceptions are in the early 1980s and in 2009, following the financial crisis. \\
We notice in the data also the effect of the slow transitioning towards renewable sources with a most significant increase over the last years for this type of energy. However, the growing energy consumption makes more difficult the challenge of transitioning the energy systems away from fossil fuels towards low-carbon sources of energy. New low-carbon energy has indeed to meet this additional demand and to try to displace existing fossil fuels in the energy mix. \\
We exploit more this topic by reporting in Figure \ref{cons_totyear} the rankings based on energy consumes for sectors and countries. In particular, we analyse the ranking with respect to both years and types of energies. Focusing on countries (see Figure \ref{cons_coutype}), we observe a high stability in the rankings based on non-renewable energy sources. Although the presence of most relevant economies (as United States, China, Japan and Russia) is confirmed in all the periods, we notice the increasing role of Asian and African countries (as South Africa, India, Taiwan and South Korea) with respect to European countries (as France, Ukraine and Great Britain) and Canada. \\
Considering instead renewable sources, except for China and United States, that are characterised again by a very high level of consumes, we observe Brazil, Canada, India, Indonesia and South Africa in the top positions. It is also noticeable the role of Nordic countries (as Norway and Sweden), that although based on a lower absolute value of consumes due to the size of country and population, showed the top positions at early times of energy transition.

\begin{figure}[htbp]
	\begin{subfigure}[t]{1\textwidth}
		\includegraphics[scale=0.25]{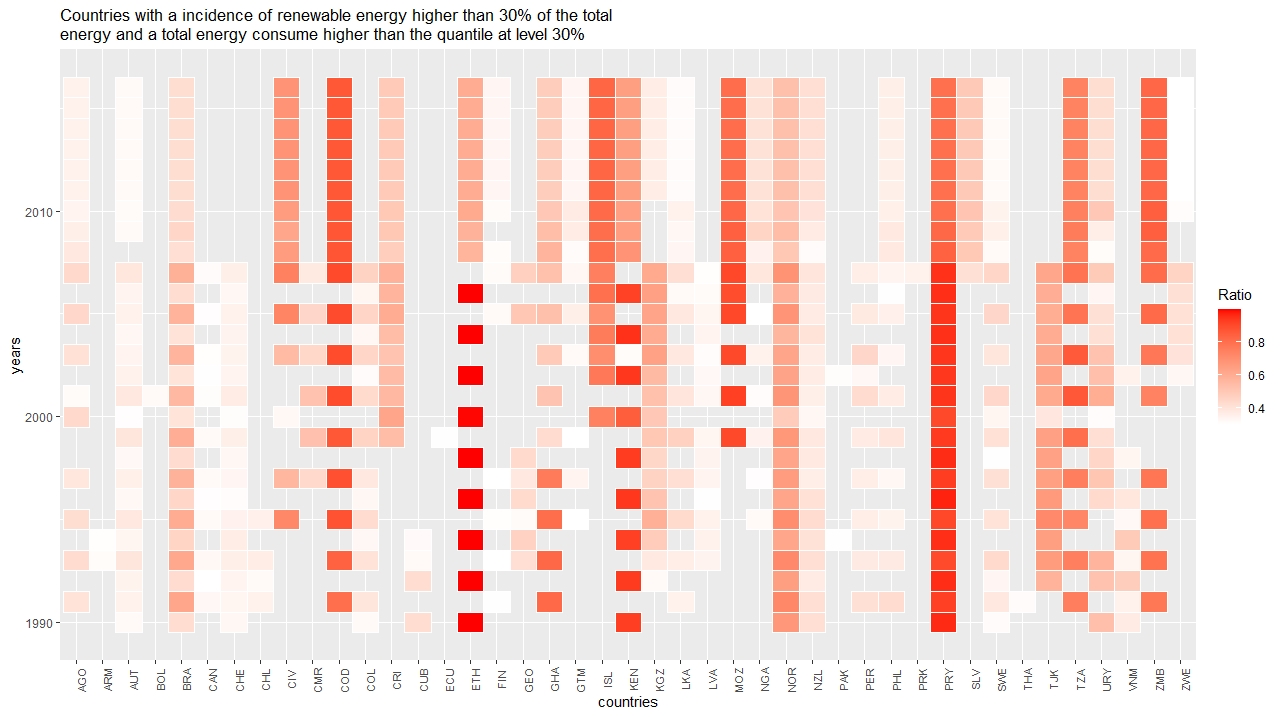}
		\caption{Countries with an incidence of consumes due to renewable energy sources higher than 30\% of total consumes. We display only countries and periods in which the energy consume is higher than the quantile of the distribution at level 30\%. Data are reported for all the analysed period.}
		\label{PercEn_country}
	\end{subfigure} \\
	\begin{subfigure}[t]{1\textwidth}
		\includegraphics[scale=0.25]{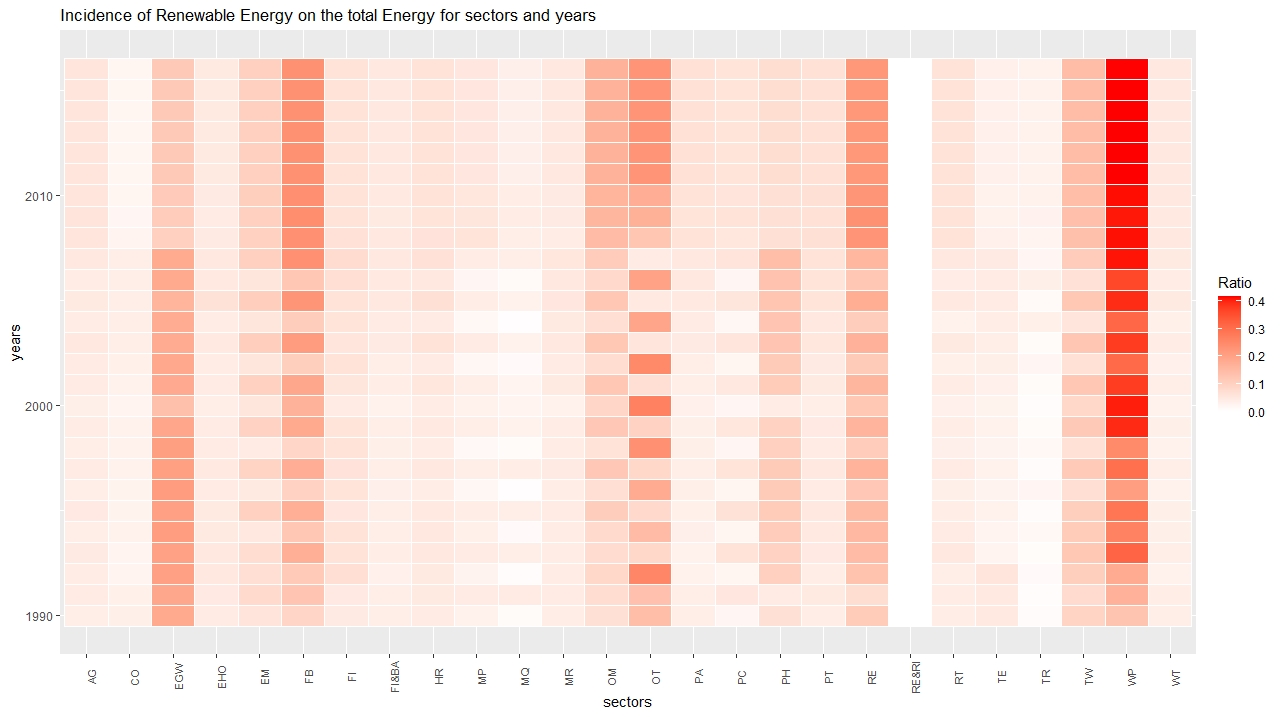}
		\caption{Incidence of consumes due to renewable energy with respect to the total consumes for each sector.}
		\label{PercEn_sector}
	\end{subfigure} 
	\caption{Proportion of consumes due to renewable energy sources }
	\label{PercEn}
\end{figure}

In order to better analyse how countries and sectors are transitioning towards new energy sources, we computed the incidence between consumes due to renewable energy and total energy consumes (see Figure	\ref{PercEn}). In particular it is noticeable the presence of several countries that are leading the change on renewable energy. Indeed, the need for a swift transition to renewable energy is more urgent than ever and there are several countries that are using clever combinations of renewable resources and targeted policies to drive down their emissions. \\
Focusing on Figure \ref{PercEn_country}, we concentrate on countries that has moved part of their consumes toward renewable sources. In particular we notice some peculiar patterns with different macro-areas characterised by a significant development of renewable energy usage. \\
It is confirmed the large diffusion of renewable sources in Nordic Europe (see, also, \cite{NER}). In particular, Sweden has reached a very high target of renewable energies by taking advantage of their natural resources and using a combination of hydroelectric power and bioenergy. Indeed, with plenty of moving water and a large forest cover, these energies became the largest renewable power sources, helping support a local energy boom. 
In Norway, the electricity production comes almost totally from renewable energy sources. Hydropower has been indeed the basis for Norwegian industry and the development of a welfare society since the country started utilizing the energy in rivers and waterfalls to produce energy in the late 1800s. 
A bit lower, but characterised by a well development, is the incidence of renewable consumes in Finland.  In this case, the most important forms of renewable energy are bioenergy; fuels from forest industry side streams and other wood-based fuels, hydropower, wind power and ground heat. Bioenergy is also generated from biodegradable waste and side streams of agriculture and industrial production. A very high ratio has been observed for Iceland. Indeed the country is a world leader in renewable energy. The electricity grid is produced only from renewable resources. In terms of total energy, 85\% of the total primary energy supply in Iceland is derived from domestically produced renewable energy sources. \\
South America is one of the world's leading renewable energy-producing regions (see, also, \cite{IRENA}). In particular Brazil, Venezuela and Colombia are nowadays larger producer in that area of the world. However, it is noticeable the presence of several country with a very high energy efficiency. In a time where global emphasis is on decarbonization and emission reduction and countries are discussing climate change, Paraguay is taking advantage of the potential of being a country with huge renewable energy resources, reaching an incidence of 80\% with respect to total consumes. Additionally, Paraguay is the country with the cleanest electric power production in the world, as its electricity generation has zero carbon dioxide emissions, according to the data from the World Economic Forum (\cite{WEF}) and it is among the countries with the highest production of hydroelectricity per capita at the global level. \\
Also Uruguay has traditionally been based on domestic hydroelectric power along with thermal power plants, and the it relies on imports from Argentina and Brazil at times of peak demand. Over the last years, investments in renewable energy sources, such as wind power and solar power, allowed the country to cover the greatest part of its electricity needs with renewable energy sources and to become also an exporter. Other examples can be represented by Colombia, Chile, Costa Rica and Ecuador that are using combinations of various renewable energy sources to cover their needs. \\
Africa has vast resource potential in wind, solar, hydro, and geothermal energies and falling costs are increasingly bringing renewables within reach. Top five greenest African countries are well represented in Figure \ref{PercEn_country}. In particular, using more than 90\% of green and sustainable energy, Mozambique is one of the greenest countries in the world, with hydroelectric power accounting for the vast majority of its installed capacity and the remainder being made up of solar, gas, and other sources. With an abundance of hydropower and solar, also Ethiopia cover more than 60\% of the consumes with renewable energy. Kenya derives most of its energy from geothermal sources and it is one of the world leaders in renewable energies. In 2016, the east-African country generated 26\% of Africa's entire renewable energy in terms of wattage and is a global leader in its number of solar power systems installed per capita. Ghana has one of the highest rates of access to electricity in Africa, with its power supply consisting mainly of hydroelectricity and thermal sources, and the remaining due to solar power. Zambia has largely relied on hydroelectric power and diesel generation produced by a state-owned company that operates the electricity market. Over the last five years, the country has seen an increase in the use of coal and heavy fuel oil to produce electricity. Finally, a peculiar situation is observed in the Democratic Republic of Congo. This country has an immense and varied energy potential, consisting of non-renewable resources, including oil, natural gas, and uranium, as well as renewable energy sources, including hydroelectric, biomass, solar, and geothermal power. On the one hand, a very high incidence of renewable sources is observed, thanks to the largest potential in Africa and one of the largest worldwide. On the other hand, the country is mired in energy poverty. Indeed, according to a report published in 2019 (see \cite{PowerAfrica}), only a small part of the population has access to the electricity (10\% in rural area and 40\% in urban areas). \\
Focusing on Figure \ref{PercEn_sector}, we notice that sectors, whose transition is favoured, have a higher incidence of renewable consumes. On average, lower ratios are observed in the industrial sectors and in transports. Indeed, although renewable energy has received a good deal of attention for power generation and for residential applications, its use in industry has attracted much less attention (see \cite{UN}). However, there is an important exception. Indeed, in Figure \ref{PercEn_sector}, while all the ratio are lower than 25\%, Wood and Paper (WP) sector shows an incidence higher than 40\%. Paper industries are indeed larger energy consumers for which it is easier to transition to renewable sources because almost every wood product can be recycled and used to generate clean and carbon-neutral energy.

\subsection{Nodes and Layers centrality}\label{sec:centr}

We focus now on the results of the application of the MD-HITS approach described in Section \ref{sec:n2} to the embodied energy flow in the temporal multilayer network $\mathcal{G}$. The approach has been applied considering all the years involved obtaining, in this way, two scores for each layer (hub and authority) and two scores for each node (broadcasting and receiving) that are based on the assessment of the relevance of sectors and economies in the whole period. \\

\begin{figure}[H]
	\includegraphics[scale=0.25]{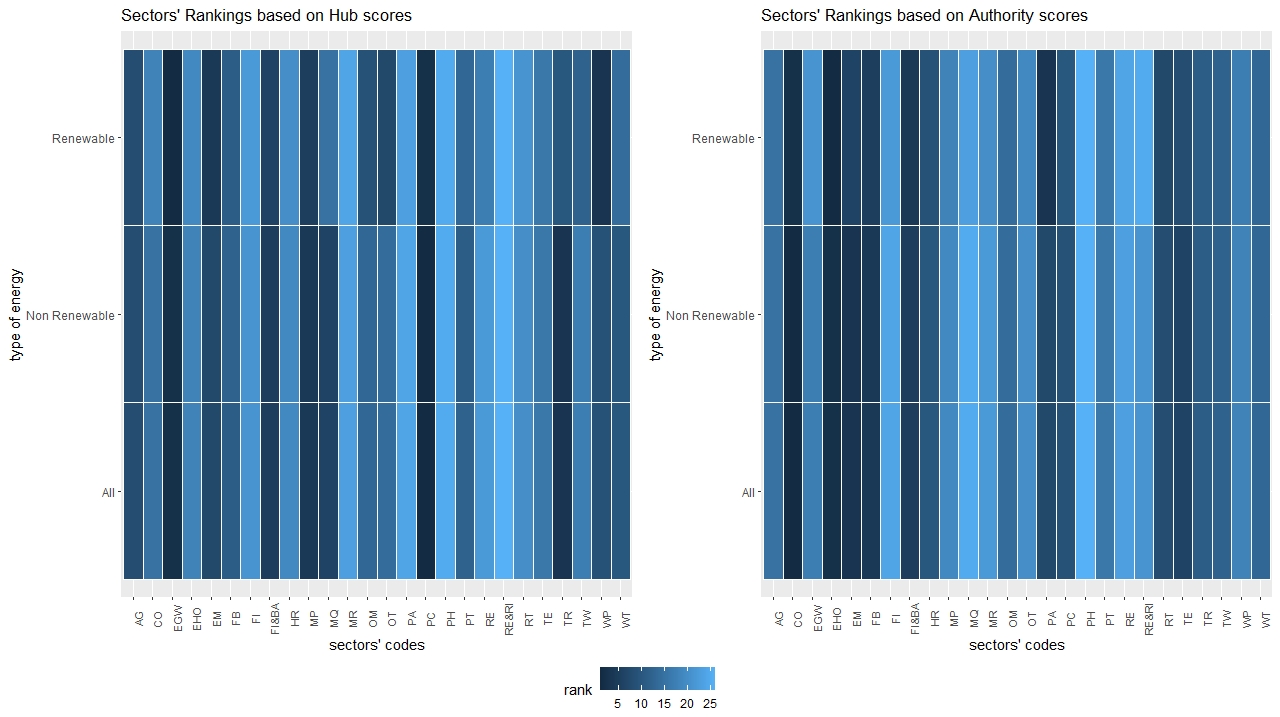}
	\caption{Ranking of sectors based on hub and authority scores. The ranking has been obtained by applying the MD-HITS procedure on the temporal multilayer and considering the three alternative networks based on different energy types.}
	\label{rank_sectortype}
\end{figure}

To this end, we display in Figure \ref{rank_sectortype} the hub and authority rankings of the different sectors based on the three alternative types of energy (i.e. renewables, non-renewables and all types). We notice a very high rank correlation between the scores based on all the energy sources and the non-renewable ones. This is mainly due to the higher relevance of non-renewable energy flows on the total consumes. \\
Focusing on hubs we have that specific sectors play a relevant role both when renewable or non-renewable sources are considered. In particular, Electricity (EGW), the Petroleum, Chemical and Non-Metallic Mineral Products (PC), Metal Products (MP) and Financial Intermediation and Business Activities (FI\&BA) belong to the top six in both cases. Those sectors are indeed among the major contributors to the global economic development (see, e.g., \cite{EUenergy}). \\
We have instead that other sectors as Transport (TR), Mining and Quarrying (MQ) and Wood and Paper (WP) show a different pattern according to the energy source considered. In particular, the first two sectors show a very high ranking when non-renewable sources are considered, while Wood and Paper (WP) is in the top ranking based on renewable energy. This behaviour can be justified by the fact that TR and MQ are characterised by an intensive use of non-renewable source, while, as shown also in the previous section, WP industries appear as leading energy transition. \\
These results can be also interpreted in terms of pollution and can be used to provide a suitable tool for policymakers for the assessment of alternative strategies to reduce the impact on the environment (see, e.g., \cite{Zhao2020} and \cite{MADURAIELAVARASAN2022112204}). For instance, energy and manufacturing sectors produce approximately the $42\%$ of CO$_2$ emission worldwide (see, e.g., \cite{friedlingstein2010} and \cite{huisingh2015}) and are 
classified among the most polluting sectors (see, e.g., \cite{Birol2017}, \cite{canadell2007}, \cite{friedlingstein2010}, \cite{huisingh2015}). These industries are indeed involved in the production of primary (and secondary) goods and services and characterised by a strong dependence on fossil fuels in both the generation and the processing phases (see \cite{WANG2019}).\\
The analysis of sectors is completed by the authority rankings which identify the main sectors requiring energy flows to manufacture products. Authority scores appear very similar when different energy sources are taken into account.  Top ranking is composed by Construction (CO), Education, Health and Other Services (EHO), Electrical and Machinery (EM), Food and Beverages (FB). In this case, we have sectors that supply final or intangible goods. However, CO and EM industries use large volume of diesel for machinery as well as electricity for powering buildings and tools, but also present many opportunities to save energy. Therefore, proper strategies for an energy-efficient design can be developed (see, e.g. \cite{melgar2022}). \\
Furthermore, significant differences in hubs and authority rankings can be noticed. This behaviour is not surprising as industries differ for product development processes and input factors (human capital, money, etc.). In general, sectors producing intermediate or final goods are characterised by higher authority scores because the final products are at the end of the supply chain. Conversely, sectors selling primary or secondary goods are characterised by higher hub scores being, their products, at the top of the supply chain.\\

\begin{figure}[H]
	\includegraphics[scale=0.25]{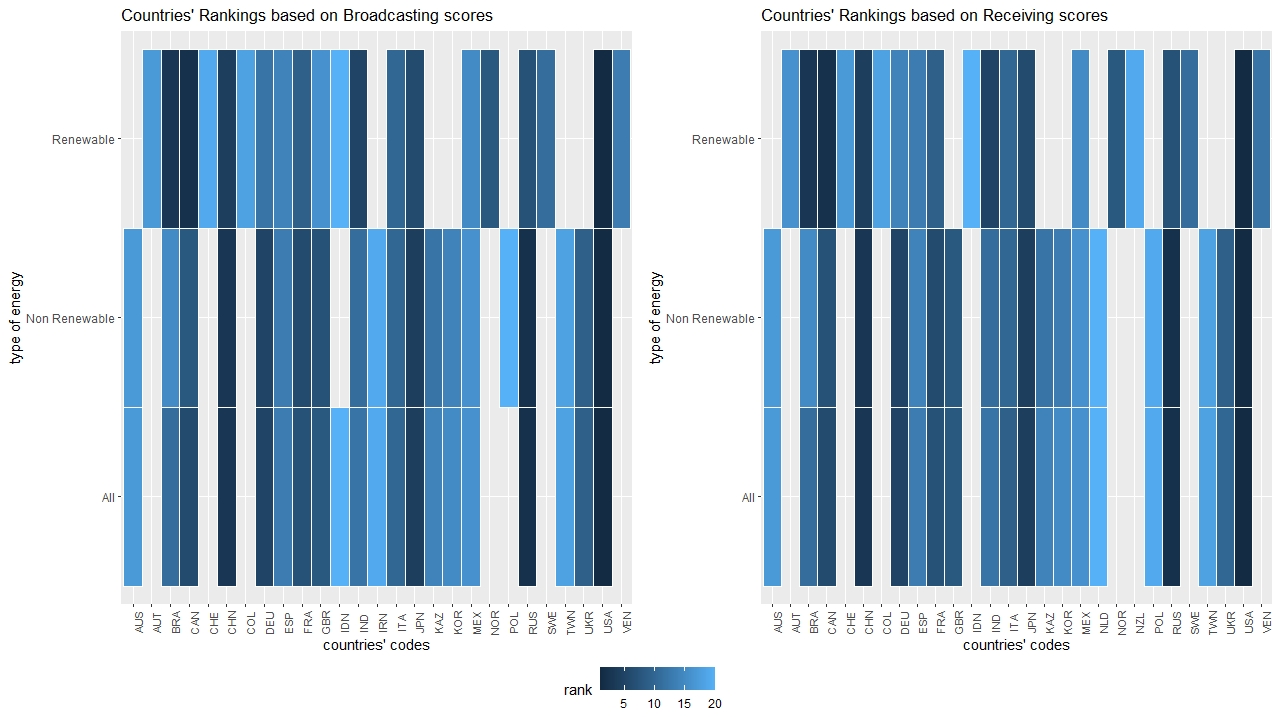}
	\caption{Ranking of countries that belong to the top 20 based on broadcasting and receiving scores obtained by applied the MD-HITS procedure on the temporal multilayer. We display only countries that belong to the top 20 in at least one year. The approach has been repeated considering the three alternative networks based on different energy types.}
	\label{rank_coutype}
\end{figure}

As regards to countries, we focus in Figure  \ref{rank_coutype} on broadcasting and receiving centrality rankings. In particular, we display only countries that belong to the top 20 in the rankings based on at least one energy source. Unlike the sectors, we have a high correlation between broadcasting and receiving scores when the same energy source is considered. Interesting patterns emerge when different energies are analysed. 
Focusing on non-renewable sources, it is prominent the relevance of countries with a high internal consumption and, on average, with high levels of international trades in the period, as United States, Russia, China, Japan and Germany. The results emphasize the central positions of these countries in the commercial trade but also in terms of energy consumption. Hence, specific attention must be paid also to their contribution in terms of environmental damages. In fact, international trade consists mainly of tangible goods requiring fossil fuel-powered procedures to be produced and exported (see \cite{huisingh2015}).  \\
Additionally, some countries trade less extensively than others, due to a large domestic market and a limited number of connections with neighbouring regions. In particular, several European countries (as Italy and Germany) and Eastern Asian countries (as China, India, Japan and South Korea) have a significant negative balance trade, while Russia and Middle East (e.g., Saudi Arabia and Emirates) provide an opposite behaviour being important exporters\footnote{see World Energy and Climate Statistics for details \url{https://yearbook.enerdata.net/total-energy/world-import-export-statistics.html}}. In general, since the network considers non-renewable embodied energy flows and it is affected by both consumes and trades in the domestic and in the international markets, we observe that most central countries have a relevant role both in terms of consumes and demands.\\
We focus now on the results based on renewable sources. Although dominant countries in the international trade (as United States, China and Russia) belong again to the top positions, we observe the emergence of leading countries in the transition towards renewable energy. Brazil, Canada, Norway and Sweden are indeed in the top ten and Venezuela, Colombia and New Zealand follow closely. In particular, we can appreciate the presence of Nordic countries that have topped the world in terms of Energy Transition Index (see \cite{WEFETI}) because of their energy performance, including the resilience and the efficiency of generation and transmission, and their progress to cleaner forms of energy. Similarly, we have countries whose economy is characterised by a large production of renewable energy. Canada and Brazil are indeed the second and the third largest producer of hydroelectricity in the world\footnote{see IRENA - International Renewable Energy Agency data and statistics.}. 

\begin{figure}[H]
	\begin{subfigure}[t]{1\textwidth}
		\includegraphics[width=12cm,height=8cm]{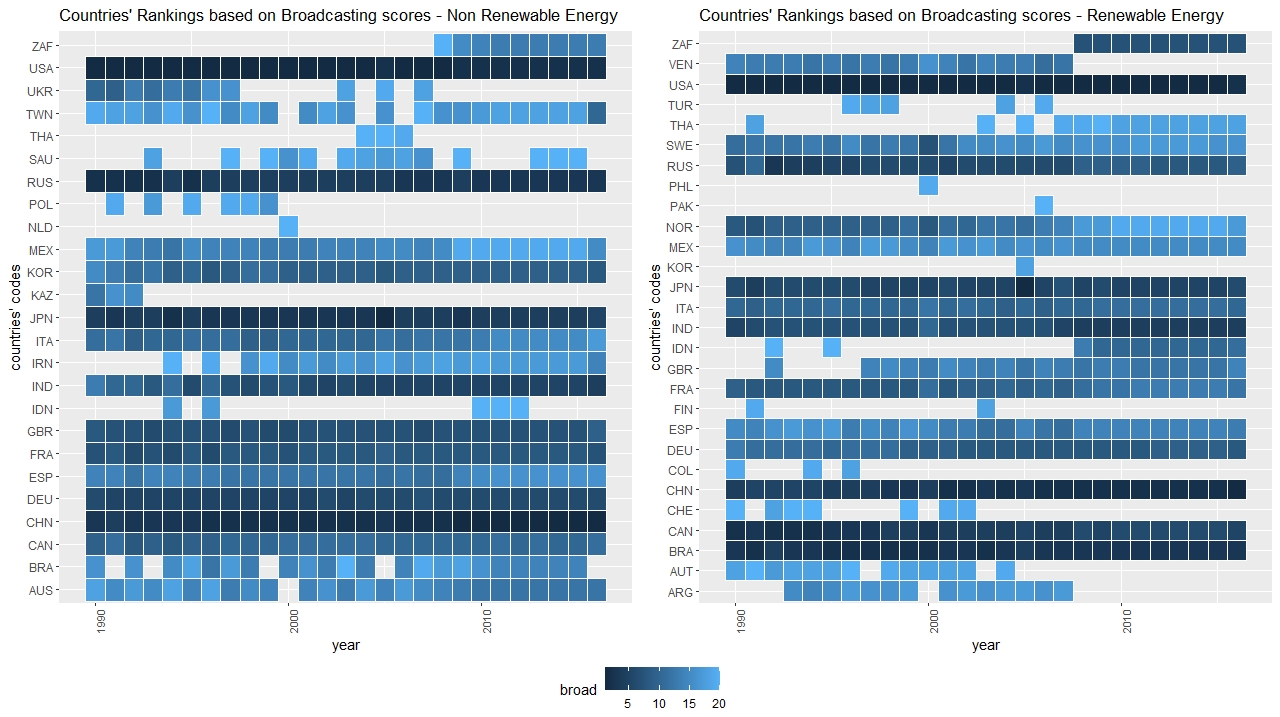}
		\caption{Ranking of countries over time based on broadcasting scores in case of non-renewable and renewable energy. We display only countries that belong to the top 20 in at least one year.}
		\label{Temp_countrybroad}
	\end{subfigure} \\
	\begin{subfigure}[t]{1\textwidth}
		\includegraphics[width=12cm,height=8cm]{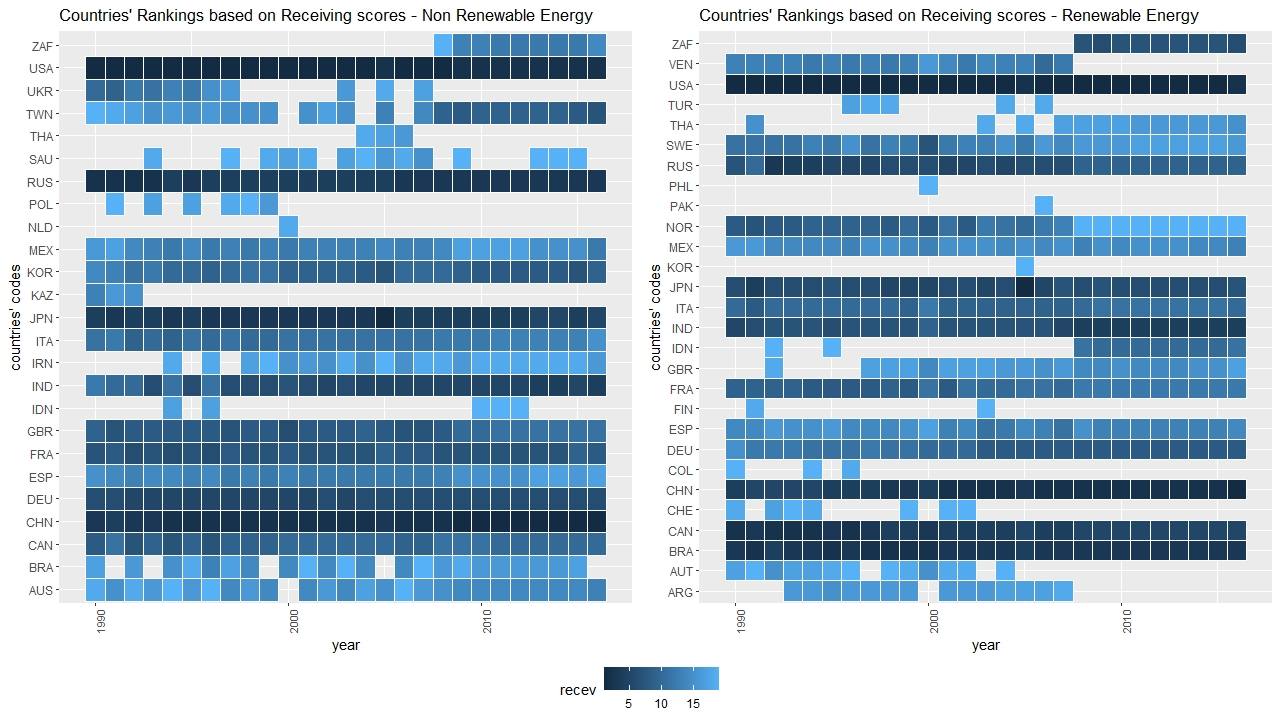}
		\caption{Ranking of countries over time based on receiving scores in case of non-renewable and renewable energy. We display only countries that belong to the top 20 in at least one year.}
		\label{Temp_countryrecev}
	\end{subfigure} 
	\caption{Ranking of countries over time.}
	\label{Temp_country}
\end{figure}

The MD-HITS have been also applied separately for each year in order to analyse the evolution over time of sectors and countries' centralities. In this case, the approach, described in Section \ref{sec:n2}, has been implemented separately considering each network $\mathcal{G}_{t}$. This is a particular case of the approach presented in Section \ref{sec:n2}, when we consider only one time period, i.e., $t=1$. In this way, for each period we obtain two scores for nodes and two scores for layers and we can appreciate how countries' and sectors' rankings evolve.
According to sectors\footnote{For the sake of brevity we do not report the Figures with detailed results.}, a very stable behaviour is obtained over time with patterns very close to the ones displayed in Figure \ref{rank_sectortype}. \\
We focus instead in Figure \ref{Temp_country} on the evolution of the broadcasting and receiving scores. As for the previous analyses, we distinguish the results according to the type of energy sources. Also when each time period is considered separately, it is confirmed a very high correlation between broadcasting and receiving scores of countries when the same energy is taken into account. \\
Focusing on non-renewable sources, we observe no significant changes in the top positions, but it is interesting the raise of new countries. In particular, we notice the presence of Iran and Saudi Arabia, that emerge over recent years in Figure \ref{Temp_country} (left side). These countries are indeed leading exporters and consumers of fossil fuels, while renewable energy technologies currently do not have a significant and adequate role in the energy supply (\cite{Solaymani} and \cite{Barhoumi}).
It is also noticeable the case of South Africa that since 2008 belongs to the top position being highly dependent on coal (see, e.g., \cite{AKINBAMI}). However, since the country is characterised by good sunshine and coastlines that lend themselves to wind power generation, the transitioning toward new sources leaded South Africa in the top positions also when renewable energy sources are considered (see Figure \ref{Temp_country} - right side). \\
As regards to renewable energy, it is confirmed the presence of Nordic and American countries, as in Figure \ref{rank_coutype} when the whole period is considered.  However, in Figure \ref{Temp_country} it could be also noticed the increasing attention towards renewable sources of biggest players in the European market (as France, Germany, Great Britain, Italy and  Spain) because of their relevance in terms of trades and consumes. Over the last years, it is worth mentioning also the raising role of Thailand and Indonesia.
Renewable energy is indeed a sector that is developing in Thailand. With its current level of carbon emissions, the Thai government is following neighbourhood countries by promoting renewable energy to cut the dependency on fossil fuel imports, especially natural gas, and reduce the environmental impact from traditional energy sources (see, e.g., \cite{Diewvilai}). The depletion of natural gas reserves and the increasing fuel import bills are becoming major challenges for this country.  \\
As part of the Paris Agreement, Indonesia has pledged to reduce its national emission within the next decade. The government committed in a new \lq\lq Global Coal to Clean Power Transition Statement\rq\rq to phasing out coal power, scaling up clean power and ensuring transition away from coal. Due to its geography, the country is naturally blessed with a massive potential of renewable energy resources on both land and sea (see \cite{Langer}). In particular, the country also controls 40\% of the globe's geothermal reserves and, therefore, energy demand raised along with the economic development and the population growth. 

\subsection{Arcs' centrality}\label{sec:ecentr}

\begin{figure}[H]
	\includegraphics[width=6.5cm,height=6cm]{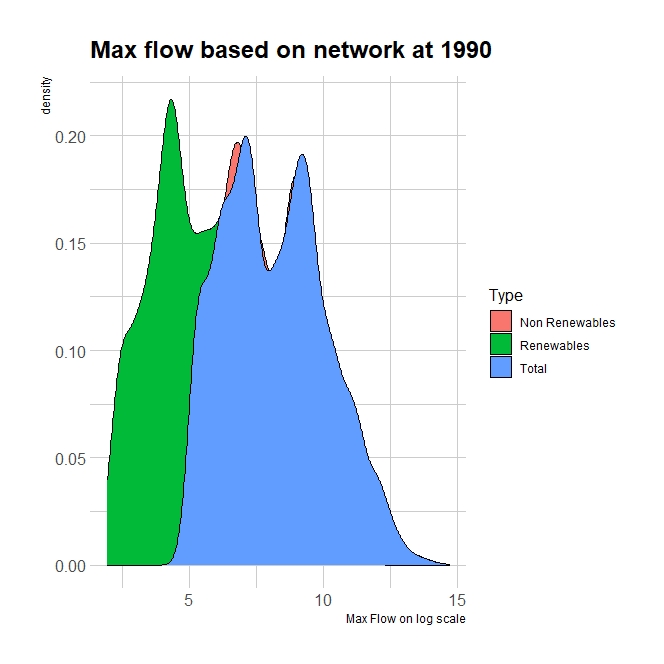}
	\includegraphics[width=6.5cm,height=6cm]{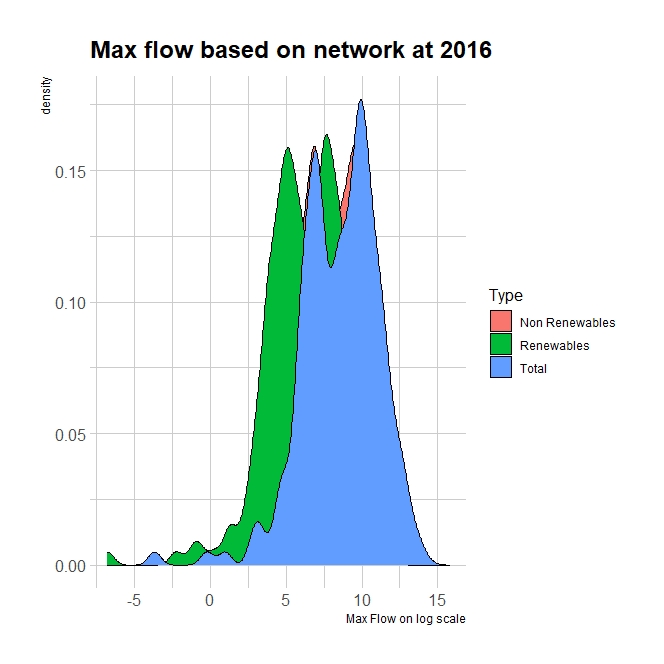}
	\caption{Distribution of max flow (reported on log scale) for two different time periods (1990 and 2016) and for three alternative types of energy (renewable, non-renewable and all sources).}
	\label{Maxflow}
\end{figure}

As described in Section \ref{sec:n3}, we focus on the identification of the most relevant connections in the network. To this end, we computed for each time period the maximum flow for each pair of vertices. In this case the analysis has been simplified by working directly at country level in order to evaluate which connections between economies have a strategic role in the network. We display in Figure \ref{Maxflow} the distributions of max flows in 1990 and 2016, respectively, obtained on the multilayer networks built using alternative energy sources. Two peculiar aspects can be noticed. On the one hand, the distinction between renewable and non-renewable is remarkable in 1990, while over time, a higher similarity between the networks based on different energies can be appreciated. On the other hand, the average max flow is increasing over time with a higher growth for renewable energies. Both effects are mainly due to the transition towards a wider use of renewable energy sources as well as to a higher density in the network.

\begin{figure}[H]
	\includegraphics[width=12cm,height=6cm]{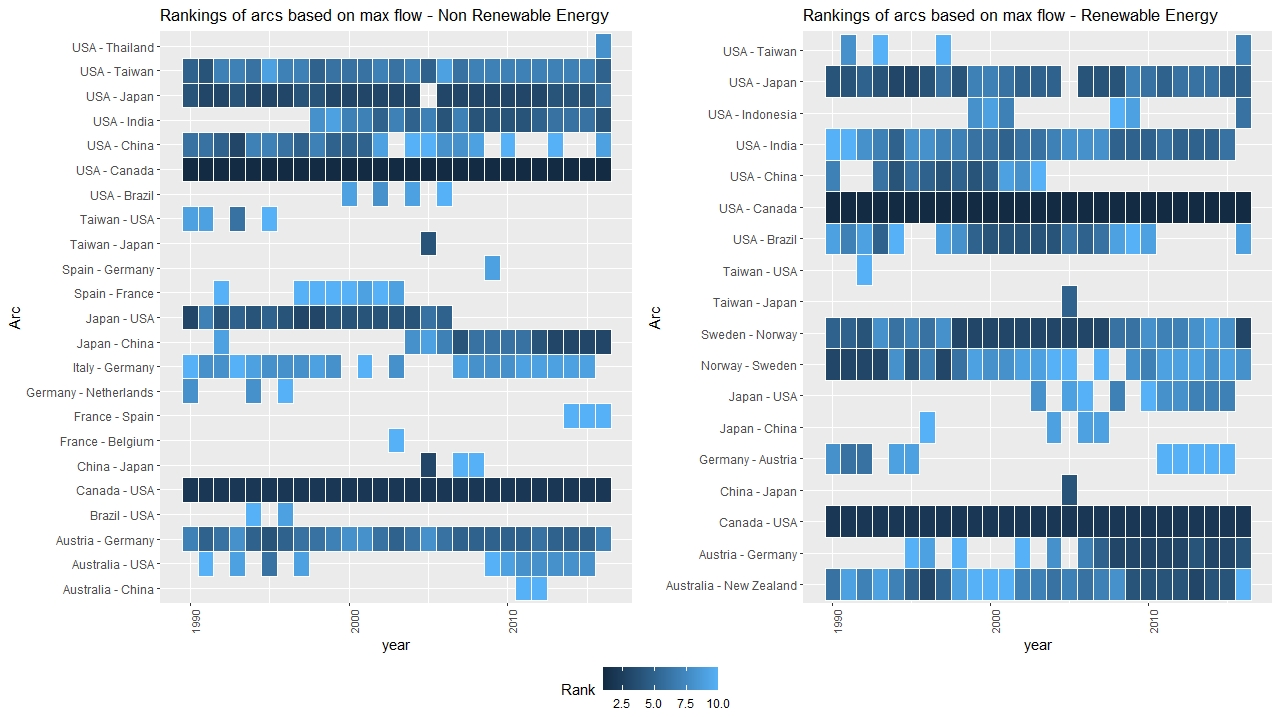}
	\caption{Ranking of most relevant arcs in terms of reduction of total max flow. We display the results divided between renewable and non-renewable resources and for different time periods. We report only arcs that belong to the top 10 in at least one year.}
	\label{MaxflowRemo}
\end{figure}

We focus now on the identification of most relevant arcs in the network. To this end, for each network, based on different energies and various time periods, we compute the max flow matrix $\widetilde{\mathbf{F}_t}$,  defined in Subsection \ref{sec:n3} removing in turn each arc. Then for each arc, we compute the ratio $\widetilde{r}_t$ defined in (\ref{coeff_r}). In this way, we identify the arc, whose removal, has the highest effect on the total max flow of the network.\\
In Figure \ref{MaxflowRemo}, most relevant arcs are displayed considering different time periods and non-renewable and renewable energies, respectively. Focusing on non-renewable sources (left side), we notice how advanced economies and geographical proximity appear as two relevant factors. We have indeed in the top ranking connections that regard United States with Canada and with other relevant players in the Asia-Pacific area (as China, Japan, India, Taiwan, Australia). Additionally, some specific European connections, mostly involving Germany, emerge. \\
As regards to the renewable case, we notice again the effect of the relevance of the economy as well as the presence of arcs involving countries that are characterised by an advanced transition towards renewable sources. Although the connections between United States and other players appear again in the top ranking, it is noticeable the presence of arcs connecting Nordic countries. Also Indonesia and New Zealand are now present in some strategic arcs.

\section{Conclusions}
\label{se:conc}
In this paper we provide a methodology for assessing relevant countries and sectors in terms of energy flows embodied in trade patterns. To this end, we consider a temporal weighted and directed multilayer network that extends classical input-output analysis. To evaluate relevant economies and industries in the network, we apply a multilayer generalisation of hub and authority centrality measures by assigning two scores to each node and two scores to each layer over the whole period. Additionally, strategic connections are identified by means of a method, new in this context, based on the effect on the maximum flow of the links' removal from the network. \\
A numerical analysis has been developed by taking into account a long time period and the data of the international trade. Results prove that the methods are effective in catching the different patterns between renewable and non-renewable energy sources. The proposed approach can be a useful tool for policymakers in order to develop an environmental
strategy that considers environmental degradation, energy availability and economic growth. Indeed, the multilayer structure, considering the energy supply chain, allows to trace the effects on the system of exogenous and endogenous shocks as, for instance, changes in countries'
environmental policies.


\bibliographystyle{Chicago}
\bibliography{Myref}   

\appendix

\newpage
\section{Maximum flow problem formulation}
\label{appsec:maxflow}
We briefly describe here the maximum flow problem which is essential for the methodology described in Sec.\ (\ref{sec:n3}) to compute the total maximum flow circulating in the network.

We consider a directed graph $G = (V,E,w)$ where $V$ and $E$ are the set of nodes and arcs, respectively and $w$ is a non-negative real value associated with each arc, representing the capacity of the arc. 
In particular, the capacity $w_{ij}$ represents the maximum amount of flow that can pass from node $i$ to $j$ through the arc $(i,j)$. 


In the set $V$, we select two nodes that we label as source and target node. We denote them by $n_s$ and $n_t$, respectively. 
The source node $n_s$ sends some units of flow to the target node $n_t$ through the arcs of the graph $G$. We intend to determine the maximum flow sent from $n_s$ to $n_t$ through an arc (if source and target nodes are adjacent) or a pattern connecting $n_s$ to $n_t$. 
This problem can be solved by using a constrained optimization model, the so-called max flow problem (see \cite{kleinberg2006}), that ships the maximum possible amount of flow from the source node to the target one respecting the constraints.
We denote by $f_{ij}$ the unit of flows passing through the arc $(i,j)$.
The objective function to be maximised represents the maximum amount of flow that passes from a source node to a target one, and the pattern of such maximum flow is determined by choosing among the arcs in $E$. Denoting this function as $|f|$, we have:
$$
|f| = \sum_{n:(n_s,n)\in E} f_{nn_s}.
$$

The constraints deal with the problem of admissible flow. At first, it is clear that the variable $f_{ij}$ cannot exceed the arc capacity $w_{ij}$. We have to take into account these conditions in searching the max flow for each arc $(i,j)$ belonging to the pattern from $n_s$ to $n_t$. This leads to the following constraints:

\begin{equation*}
	0 \le f_{ij} \le c_{ij} \hspace{3mm} \forall (i,j)\in E.
	\label{upper_bound}
\end{equation*}

Another condition, quite obvious, is that the sum of the flows entering in a node must be equal to the sum of the flows exiting from the node, except for the source and the target nodes. This condition expresses the  law of conservation flow:
\begin{equation*}
	\sum_{i \in V: (i,n) \in E} f_{in} - \sum_{i\in V: (n,i) \in E} f_{ni} = 0 \hspace{2mm} \forall n\in V\setminus \{n_s,n_t\}.
	\label{conservation_flow}
\end{equation*}
\\

Then, the maximum flow problem can be stated as follows
\begin{eqnarray*}
	\max_{f_{ij}} & |f| & \\
	\text{s.t.} & \nonumber\\
	&\sum\limits_{i \in V: (i,n) \in E} f_{in}-&\sum\limits_{i\in V: (n,i) \in E} f_{ni}=0 \hspace{2mm} \forall n\in V\setminus \{n_s,n_t\} \\
	& 0 \le f_{ij} \le c_{ij} & \forall (i,j)\in E.
\end{eqnarray*}

\newpage
\section{List of sectors and countries}
\label{sectorscountrieslists}

\begin{table}[!h]
	\begin{tabular}{lc}
		Sectors	& Code \\
		\hline \hline
		Agriculture	& AG \\
		Fishing	& 	FI\\
		Mining and Quarrying	& 	MQ\\
		Food and Beverages 	& 	FB\\
		Textiles and Wearing Apparel 	& 	TW\\
		Wood and Paper 	& 	WP\\
		Petroleum, Chemical and Non-Metallic Mineral Products	& 	PC\\
		Metal Products 	& 	MP\\
		Electrical and Machinery 	& 	EM\\
		Transport Equipment		& TE\\
		Other Manufacturing 	& 	OM\\
		Recycling 	& 	RE\\
		Electricity, Gas and Water	& 	EGW\\
		Construction	& 	CO\\
		Maintenance and Repair	& 	MR\\
		Wholesale Trade 	& 	WT\\
		Retail Trade	& 	RT\\
		Hotels and Restaurants		& HR\\
		Transport	& 	TR\\
		Post and Telecommunications	& 	PT\\
		Financial Intermediation and Business Activities 	& 	FI\&BA\\
		Public Administration 	& 	PA\\
		Education, Health and Other Services	& 	EHO\\
		Private Households 	& 	PH\\
		Others		& OT\\
		Re-export and Re-import 	& 	RE\&RI\\
		\hline
	\end{tabular}
	\caption{List of sectors.}
	\label{listsectors}
\end{table}

\newpage
\footnotesize
\begin{longtable}{lccccc}
	\hline
	\textbf{Country} & \textbf{Code} & \textbf{Country} & \textbf{Code} & \textbf{Country} & \textbf{Code}\\
	\hline
	Afghanistan & AFG & Albania & ALB & Algeria & DZA \\ 
	Andorra & AND & Angola & AGO & Antigua & ATG \\
	Argentina & ARG & Armenia & ARM & Aruba & ABW \\
	Australia & AUS & Austria & AUT & Azerbaijan & AZE \\
	Bahamas & BHS & Bahrain & BHR & Bangladesh & BGD \\
	Barbados & BRB & Belarus & BLR & Belgium & BEL \\
	Belize & BLZ & Benin & BEN & Bermuda & BMU \\
	Bhutan & BTN & Bolivia & BOL & Bosnia Herzegovina & BHI \\
	Botswana & BWA & Brazil & BRA & British Virgin Islands & VGB\\
	Brunei & BRN & Bulgaria & BGR & Burkina Faso & BFA \\
	Burundi & BDI & Cambodia & KHM & Cameroon & CMR \\
	Canada & CAN & Cape Verde & CPV & Cayman Islands & CYM \\
	Central African Republic & CAF & Chad & TCD & Chile & CHL \\ China & CHN & Colombia & COL & Congo & COG \\
	Costa Rica & CRI & Cote d'Ivoire & CIV & Croatia & HRV \\ 
	Cuba & CUB & Cyprus & CYP & Czech Republic & CZE  \\
	Denmark & DNK & Djibouti & DJI & Dominican Republic & DOM \\ 
	Democratic Republic of Congo & COD & Ecuador & ECU & Egypt & EGY \\
	El Salvador & SLV & Eritrea & ERI & Estonia & EST \\ Ethiopia & ETH & Fiji & FJI & Finland & FIN \\
	Former USSR & USR & France & FRA & French Polynesia & PYF \\ Gabon & GAB & Gambia & GMB & Gaza Strip & PSE \\
	Georgia & GEO & Germany & DEU & Ghana & GHA \\ 
	Greece & GRC & Greenland & GRL & Guatemala & GTM \\
	Guinea & GIN & Guyana & GUY & Haiti & HTI \\
	Honduras & HND & Hong Kong & HKG & Hungary & HUN \\
	Iceland & ISL & India & IND & Indonesia & IDN \\
	Iran & IRN & Iraq & IRQ & Ireland & IRL \\
	Israel & ISR & Italy & ITA & Jamaica & JAM \\ 
	Japan & JPN & Jordan & JOR & Kazakhstan & KAZ \\
	Kenya & KEN & Kuwait & KWT & Kyrgyzstan & KGZ \\
	Laos & LAO & Latvia & LVA & Lebanon & LBN \\
	Lesotho & LSO & Liberia & LBR & Libya & LBY \\
	Liechtenstein & LIE & Lithuania & LTU & Luxembourg & LUX \\
	Macao & MAC & Madagascar & MDG & Malawi & MWI \\
	Malaysia & MYS & Maldives & MDV & Mali & MLI \\
	Malta & MLT & Mauritania & MRT & Mauritius & MUS \\
	Mexico & MEX & Moldova & MDA & Monaco & MCO \\
	Mongolia & MNG & Montenegro & MNE & Morocco & MAR \\
	Mozambique & MOZ & Myanmar & MMR & Namibia & NAM \\
	Nepal & NPL & Netherlands & NLD & Netherlands Antilles & ANT \\
	New Caledonia & NCL & New Zealand & NZL & Nicaragua & NIC \\
	Niger & NER & Nigeria & NGA & North Korea & PRK \\
	Norway & NOR & Oman & OMN & Pakistan & PAK \\
	Panama & PAN & Papua New Guinea & PNG & Paraguay & PRY \\
	Peru & PER & Philippines & PHL & Poland & POL \\
	Portugal & PRT & Qatar & QAT & Romania & ROU \\
	Russia & RUS & Rwanda & RWA & Samoa & WSM \\
	San Marino & SMR & Sao Tome and Principe & STP & Saudi Arabia & SAU \\
	Senegal & SEN & Serbia & SRB & Seychelles & SYC \\
	Sierra Leone & SLE & Singapore & SGP & Slovakia & SVK \\
	Slovenia & SVN & Somalia & SOM & South Africa & ZAF \\
	South Korea & KOR & South Sudan & SSD & Spain & ESP \\
	Sri Lanka & LKA & Sudan & SDN & Suriname & SUR \\
	Swaziland & SWZ & Sweden & SWE & Switzerland & CHE \\
	Syria & SYR & Taiwan & TWN & Tajikistan & TJK \\
	Tanzania & TZA & TFYR Macedonia & MKD & Thailand & THA \\
	Togo & TGO & Trinidad and Tobago & TTO & Tunisia & TUN \\
	Turkey & TUR & Turkmenistan & TKM & United Arab Emirates & ARE \\
	Uganda & UGA & United Kingdom & GBR & Ukraine & UKR \\
	Uruguay & URY & United States & United States & Uzbekistan & UZB \\
	Vanuatu & VUT & Venezuela & VEN & Vietnam & VNM \\
	Yemen & YEM & Zambia & ZMB & Zimbabwe & ZWE\\
	\hline
	\caption{List of countries}
	\label{tab:tableofcountries_1}
\end{longtable}

\end{document}